\documentclass[10pt,journal]{IEEEtran}

\usepackage[pdftex]{graphicx}
\usepackage{latexsym,amssymb,amsmath,ifthen,cite,url}
\usepackage{epic}

\newcommand{\iftwocolumn}[2]{\ifthenelse{\boolean{@twocolumn}}{#1}{#2}}

\newcommand{\ignore}[1]{}

\newcommand{\Fig}[1]{Fig.~\ref{#1}}

\newcommand{\eqdef}{\stackrel{\scriptscriptstyle\bigtriangleup}{=} }
\newcommand{\andor}{\,/\,}
\newcommand{\cond}{\hspace{0.02em}|\hspace{0.08em}}

\newcommand{\T}{\mathsf{T}}
\renewcommand{\H}{\mathsf{H}}

\newcommand{\R}{\mathbb{R}}

\newcommand{\ccj}[1]{\overline{#1}}

\newcommand{\EE}[1]{\mathrm{E}\!\left[{#1}\right]}

\newcommand{\argmin}{\operatornamewithlimits{argmin}}
\newcommand{\argmax}{\operatornamewithlimits{argmax}}

\newcommand{\msgf}[2]{\protect\overrightarrow{#1}_{\!#2}}
\newcommand{\msgb}[2]{\protect\overleftarrow{#1}_{\!#2}}

\newcommand\relphantom[1]{\mathrel{\phantom{#1}}}

\newcounter{examplecntr}
{\begin{trivlist}\small\item[]\refstepcounter{examplecntr}%
 {\bfseries Example~\theexamplecntr%
  \ifthenelse{\equal{#1}{}}{}{ (#1)}.
}}%
{\end{trivlist}}

\newcounter{theoremcntr}
\newenvironment{theorem}[1][]%
{\begin{trivlist}\item[]\refstepcounter{theoremcntr}%
{\bfseries Theorem~\thetheoremcntr%
  \ifthenelse{\equal{#1}{}}{}{ (#1)}.
}}%
{\hfill$\Box$\end{trivlist}}

\newenvironment{_proof}{\begin{trivlist}\item[]{\bfseries Proof: }
 }{\hfill$\Box$\end{trivlist}}

\newcommand{\cent}[1]{\makebox(0,0){#1}}
\newcommand{\pos}[2]{\makebox(0,0)[#1]{#2}}
\newcommand{\connectionDot}{\circle*{1}}

\newcommand{\plusSign}{%
  \begin{picture}(0,0)(0,0)
  \put(0,0){\circle{4}}
  \put(-1,0){\line(1,0){2}}
  \put(0,-1){\line(0,1){2}}
  \end{picture}
}

\newlength{\messagetablewidth}
\newcommand{\tablebox}[1]{\framebox[\messagetablewidth]{%
\begin{minipage}{0.9\messagetablewidth}#1\end{minipage}}\vspace{-0.4pt}}

\newcounter{saveequationcntr}%

\newenvironment{eqntable}%
{\begin{table}%
\setcounter{saveequationcntr}{\value{equation}}%
\setcounter{equation}{0}%
}%
{\setcounter{equation}{\value{saveequationcntr}}%
\end{table}}

\begin{document}
\DeclareGraphicsExtensions{.pdf}

\title{%
\iftwocolumn{LMMSE Estimation and Interpolation
of Continuous-Time Signals from Discrete-Time Samples Using Factor Graphs
}
{LMMSE Estimation and Interpolation\\
of Continuous-Time Signals\\ from Discrete-Time Samples\\ Using Factor Graphs%
}
}

\author{%
Lukas Bolliger,
Hans-Andrea Loeliger, 
and Christian Vogel%
\thanks{%
Lukas Bolliger and Hans-Andrea Loeliger are with the 
Dept.\ of Information Technology and Electrical Engineering, 
ETH Zurich, CH-8092 Zurich, Switzerland.
Email: \texttt{loeliger@isi.ee.ethz.ch},
\texttt{lukas@bolligernet.ch}.
}%
\thanks{%
Christian Vogel is with
the Telecomm.\ Research Center Vienna (FTW),
Donau-City-Strasse~1, A-1220 Vienna, Austria.
Email: \texttt{c.vogel@ieee.org}.
}%
\thanks{%
An abbreviated version 
of this paper was presented at 
the 2010 Information Theory \& Appl.\ Workshop (ITA), 
La Jolla, CA, Feb.~2010 \cite{BLV:ITA2010c}.}%
}

\maketitle

\begin{abstract}
The factor graph approach to discrete-time linear Gaussian state space models
is well developed.
The paper extends this approach to continuous-time linear systems\andor{}filters
that are driven by white Gaussian noise. 
By Gaussian message passing, we then obtain 
MAP\andor{}MMSE\andor{}LMMSE estimates of the input signal, or of the state, 
or of the output signal 
from noisy observations of the output signal.
These estimates may be obtained with arbitrary temporal resolution.
The proposed input signal estimation 
does not seem to have appeared in the prior Kalman filtering literature.
\end{abstract}

\section{Introduction}
\label{sec:Intro}

\begin{figure}
\newcommand{\LowPassAmplResp}{%
 \begin{picture}(0,0)(0,0)
  \put(-5,0){\vector(1,0){20}}  \put(16,2.7){\pos{tl}{$f$}}
  \put(0,-3){\vector(0,1){15}}
  \put(0,-1){\includegraphics[width=12\unitlength]{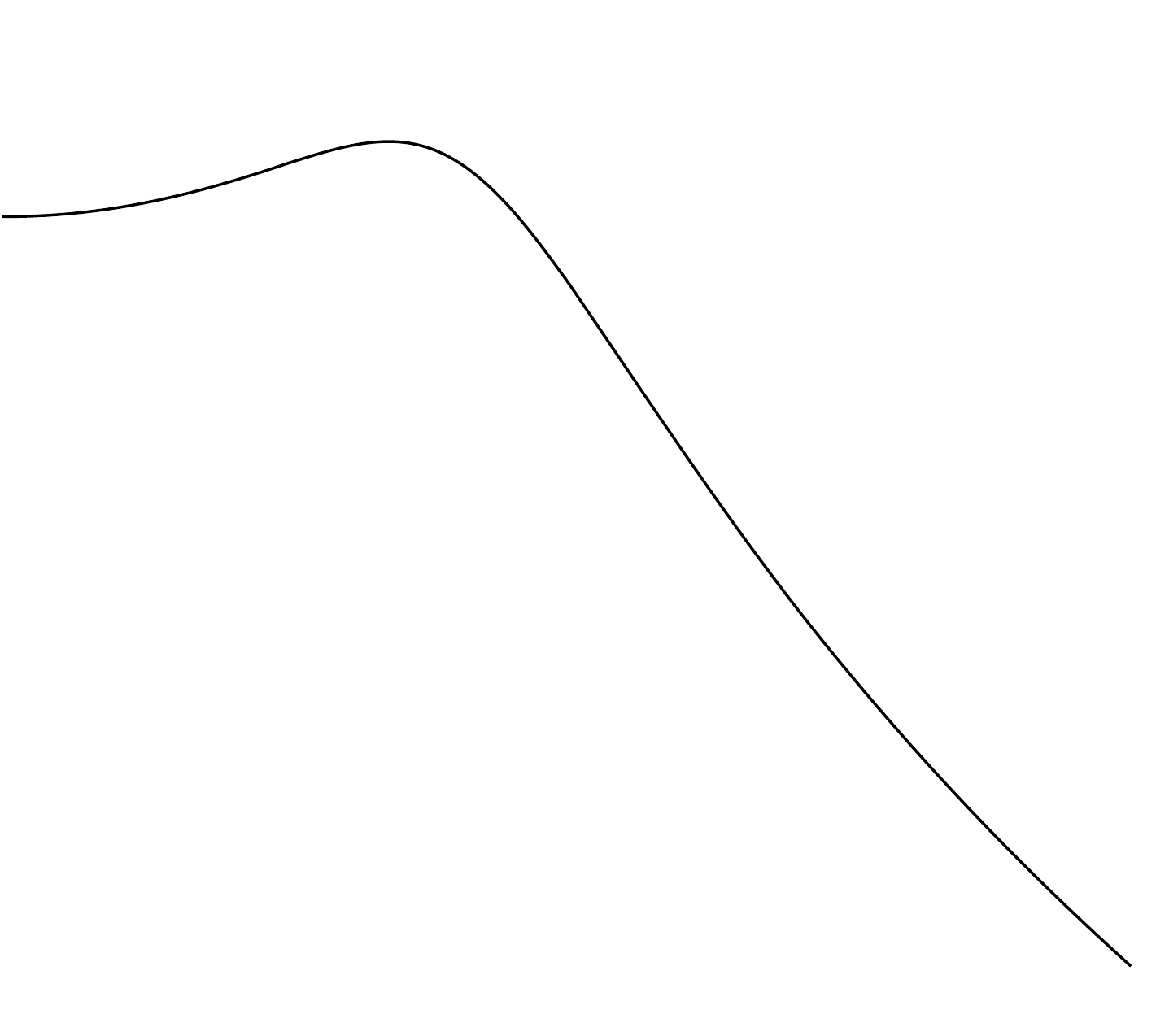}}
 \end{picture}
}
\setlength{\unitlength}{1mm}
\centering
\begin{picture}(84,32.5)(-2,-10)
%
%
\put(0,2.5){\dashbox(5,5){}}
  \put(2.5,0.5){\pos{ct}{WGN}}
  \put(2.5,-4){\pos{ct}{source}}
\put(5,5){\vector(1,0){15}}
  \put(12.5,6){\pos{bc}{$U(t)$}}
\put(20,0){\framebox(15,10){}}
  \put(27.5,-2){\pos{ct}{linear}}
  \put(27.5,-6){\pos{ct}{system\andor{}filter}}
\put(24,2.5){\setlength{\unitlength}{0.5\unitlength}\LowPassAmplResp}
\put(35,5){\line(1,0){12}}    \put(41,6){\pos{bc}{$Y(t)$}}
\put(47,5){\line(0,1){0.5}}
\put(52,5){\line(-3,2){5}}
\put(52,5){\connectionDot}
\put(52,5){\vector(1,0){11}}    \put(58,3){\pos{tc}{$Y_k$}}
\put(65,5){\plusSign}
\put(62.5,17.5){\framebox(5,5){}}
  \put(69,20){\pos{tl}{WGN}}
  \put(69,15.5){\pos{tl}{source}}
\put(65,17.5){\vector(0,-1){10.5}}
      \put(64,12){\pos{cr}{$Z_k$}}
%
\put(67,5){\vector(1,0){15}}    \put(77,3.5){\pos{ct}{$\tilde{Y}_k$}}
%
\end{picture}
\caption{\label{fig:SystemModel}%
System model.}
\end{figure}
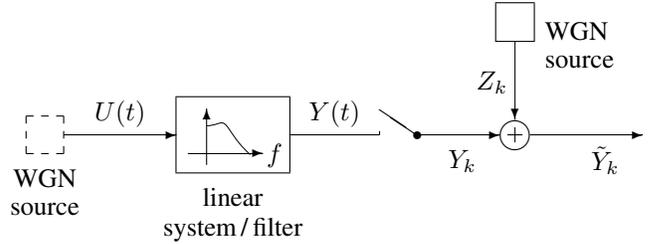

Consider the system model shown in \Fig{fig:SystemModel}: 
a continuous-time linear time-invariant system\andor{}filter is fed
by a continuous-time signal $U(t)$. 
The system output $Y(t)$ is sampled 
(at regular or irregular intervals) 
and the samples are corrupted by 
discrete-time additive white Gaussian noise. 
From the noisy samples $\tilde{Y}_k$, 
we wish to estimate the clean samples $Y_k$, or the clean signal 
$Y(t)$ at arbitrary instants $t$, or the state trajectory of the system, 
or---of particular interest in this paper---the input signal $U(t)$
at arbitrary instants $t$. 
We will not assume that any of these signals is bandlimited
(in the strict sense required by the sampling theorem); 
instead, the key assumption in this paper 
is that the given linear system has a finite-dimensional state space representation.

Problems of this kind are ubiquitous. 
For example, \Fig{fig:SystemModel} might model 
an analog-to-digital converter with a non-ideal
anti-aliasing filter and with quantization noise $Z_k$;
indeed, this application is a main motivation for this paper. 
As another example, \Fig{fig:SystemModel} 
might model a sensor with some internal dynamics 
which limits its temporal resolution of the desired quantity $U(t)$. 
In both examples,
we are primarily interested in estimating the input signal $U(t)$.

We will address these estimation problems 
under the further assumption that 
the input signal $U(t)$ is 
white Gaussian noise. 
It might perhaps seem at first that this assumption 
is problematic when $U(t)$ is actually
the signal of interest, as in the two mentioned examples. 
However, we will argue
that this assumption is meaningful in such cases and that
the LMMSE (linear minimum mean squared error) 
estimate of $U(t)$ is well defined and useful. 
An example of such an LMMSE estimate of $U(t)$ is shown in \Fig{fig:plotinpest}.
The nature of this estimate will further be illuminated
by reformulating it as a regularized least-squares problem with a penalty term $\int u(t)^2\, dt$,
as will be discussed in Section~\ref{sec:InputEstimation}.

The assumption that $U(t)$ is white Gaussian noise
turns our system model (\Fig{fig:SystemModel}) 
into a linear Gaussian model, and LMMSE estimation of 
the state trajectory or of the clean output signal $Y(t)$
amounts essentially 
to Kalman filtering%
\footnote{Note that the Kalman-Bucy filter \cite{KaBu:KalmanBucy1961}
addresses the different situation where the observations are continuous-time signals as well.}
(or rather Kalman smoothing)
\cite{Ka:1960,KaBu:KalmanBucy1961,AnMo:of1979b,MeSi:uKF1983,Haykin:AdaptFilt,GrAn:KF,KSH:LinEs2000b}.
However, estimation of the continuous-time input signal $U(t)$ 
does not seem to have been addressed in the Kalman filtering literature. 

We will also consider some extensions of the system model
including time-varying systems, vector signals, 
and systems with internal noise sources. 
These extensions are required in some of the motivating applications,
but the extensions are straightforward 
and standard in Kalman filtering.

We will address these estimation problems (as described above)
using factor graphs.
Factor graphs \cite{KFL:fg2000,Lg:ifg2004,LDHKLK:fgsp2007,Wym:ird}
and similar graphical models \cite{bi:prml,Jo:gm2004,WaJo:gmvi2008,KoFr:PGMb} allow
a unified description of system models and algorithms 
in many different fields. 
In particular, Gaussian message passing in factor graphs
subsumes discrete-time Kalman filtering and many variations of it 
\cite{KFL:fg2000,Lg:ifg2004,LDHKLK:fgsp2007,Wym:ird}.
The graphical-model approach
has facilitated the use of these techniques
as components in more general inference problems 
and it has become a mode of teaching 
discrete-time Kalman filtering
itself.

In this paper, 
we extend the factor graph approach 
to continuous-time models with discrete-time observations as described above.
This extension appears to be new%
\footnote{%
Another extension of graphical models to continuous time are 
continuous-time Bayesian networks \cite{NSK:ctbn2002c,HCFKLctbp2010,CEHFK:mfvctbn2010}, 
where the emphasis is on finite-state models and approximate inference.
Yet another such extension is \cite{VL:fgden2003c},
where linear RLC circuits are described in terms of factor graphs.%
},
and it 
significantly enlarges the domain of 
graphical models. 
We note, in particular, that
the LMMSE estimates of the continuous-time signals associated with such models 
(such as $U(t)$ and $Y(t)$ in \Fig{fig:SystemModel})
become computational objects that can be handled with arbitrary 
temporal resolution by 
Gaussian message passing. 

Applications of the methods of this paper 
(in addition to those already mentioned)
have been reported in \cite{BoLo:EUSIPCO2011c} and \cite{LBWB:ITA2011c}.
In \cite{BoLo:EUSIPCO2011c}, a new method for sampling jitter correction is proposed 
that uses the slope of $Y(t)$, which is available in the state space model, 
in an iterative algorithm. 
In \cite{LBWB:ITA2011c}, a new approach to analog-to-digital conversion 
is proposed which combines unstable analog filters with digital estimation of $U(t)$
as proposed in the present paper. 
Both of these applications build on \cite{BLV:ITA2010c} (which does not contain the proofs)
and rely on the present paper for a full justification of the proposed algorithms. 
Further applications 
(including beamforming with sensor arrays and Hilbert transforms) will be reported elsewhere.

In summary, this paper 
\begin{itemize}
\item
extends the factor graph approach to continuous-time models as in \Fig{fig:SystemModel};
\item
extends Kalman smoothing (forward-backward Gaussian message passing) 
to the estimation of input signals as $U(t)$ in \Fig{fig:SystemModel};
\item
provides the necessary background for subsequent work 
such as \cite{BoLo:EUSIPCO2011c} and \cite{LBWB:ITA2011c}.
\end{itemize}

The paper builds on,
and assumes some familiarity with,
the factor graph approach to discrete-time Kalman filtering 
as given in 
\cite{LDHKLK:fgsp2007}. 

The paper is structured as follows. 
The system model is formally stated in Section~\ref{sec:SystemModel}
and represented in factor graph notation in Section~\ref{sec:FG}.
State estimation and output signal estimation 
are then essentially obvious, 
but some pertinent comments are given
in Section~\ref{sec:GaussMessPass}.
Estimation of the input signal 
is discussed in Section~\ref{sec:InputEstimation}.
In Section~\ref{sec:NumericalExamples}, the estimation algorithms are illustrated by
some simple numerical examples.
A number of extensions of the basic system model are outlined in Section~\ref{sec:Extensions},
and Section~\ref{sec:Conclusion} concludes the paper.

The following notation will be used:
$\ccj{x}$ denotes the complex conjugate of $x$;
$A^\T$ denotes the transpose of the matrix $A$;
$A^\H\eqdef \ccj{B}^\T$ denotes the Hermitian transpose of $A$;
$I$ denotes an identity matrix;
``$\propto$'' denotes equality up to a constant scale factor;
$\mathcal{N}(m,\sigma^2)$ or $\mathcal{N}(m,V)$
denotes a normal (Gaussian) distribution with mean $m$ and variance $\sigma^2$,
or with mean vector $m$ and covariance matrix $V$, respectively.

\section{System Model}
\label{sec:SystemModel}

Let $X\in\R^n$ 
be the state of a linear system
(as, e.g., in \Fig{fig:SystemModel})
which evolves in time according to
\begin{equation} \label{eqn:ContSystDiffEq}
\dot{X}(t) = A X(t) + b\, U(t)
\end{equation}
where $\dot{X}$ denotes the derivative with respect to time 
and where both the matrix $A\in \R^{n\times n}$ and the vector $b\in \R^n$ are 
known.
The system output is the discrete-time signal 
$Y_1, Y_2, \ldots \in \R^\nu$ with
\begin{equation} \label{eqn:Yk}
Y_k = C X(t_k)
\end{equation}
where $t_1, t_2, \ldots \in \R$ (with $t_{k-1} < t_k$)
are discrete instants of time 
and where $C\in \R^{\nu\times n}$ is known. 
We will usually observe only 
the noisy output signal $\tilde{Y}_1, \tilde{Y}_2,\ldots$ 
defined by
\begin{equation}
\tilde{Y}_k = Y_k + Z_k,
\end{equation}
where $Z_1, Z_2,\ldots$ (the noise) are
independent Gaussian random variables, each of which 
takes values in $\R^\nu$ and has a diagonal covariance matrix $V_Z$.

The (real and scalar) input signal $U(t)$
will be modeled as white Gaussian noise, i.e., for $t<t'$,
the integral
\begin{equation}
\int_{t}^{t'} \!\! U(\tau)\, d\tau
\end{equation}
is a zero-mean Gaussian random variable with variance $\sigma_U^2 (t'-t)$,
and any number of such integrals are independent random variables
provided that the corresponding integration intervals are disjoint.
In consequence, it is appropriate to replace (\ref{eqn:ContSystDiffEq}) by
\begin{equation} \label{eqn:ContSystDiff}
d X(t) = A X(t)\, dt + b\, U(t)\, dt
\end{equation}
where $U(t)\, dt$
is a zero-mean Gaussian random variable with infinitesimal variance $\sigma_U^2\, dt$.

As stated in the introduction, 
we will argue later (in Section~\ref{sec:InputEstimation})
that modeling $U(t)$ as white Gaussian noise
is meaningful even when $U(t)$ is a (presumably smooth) signal of interest that we wish to estimate.

For any fixed initial state $X(t_0)=x(t_0)$,
equation~(\ref{eqn:ContSystDiff}) 
induces a probability density 
$f\big(x(t_1) \cond x(t_0)\big)$ over the possible values of $X(t_1)$
(where $t_0$ and $t_1$ are unrelated to the discrete times $\{t_k\}$ in (\ref{eqn:Yk})).
Specifically, 
integrating (\ref{eqn:ContSystDiff}) 
from $t=t_0$ to $t_1 > t_0$
yields
\begin{equation} \label{eqn:Integrated}
X(t_1) = e^{AT} X(t_0) + \int_{0}^T \! e^{A(T-\tau)} bU(t_0+\tau)\, d\tau
\end{equation}
with $T\eqdef t_1-t_0 > 0$.
If $U(t)$ is white Gaussian noise (with $\sigma_U^2$ as above), 
then the integral in (\ref{eqn:Integrated}) is a zero-mean 
Gaussian random vector with covariance matrix%
\footnote{This covariance matrix is essentially the controllability Gramian.
However, controllability is not required in this paper.}
\cite{GaWa:idm2008,So:dtss2002b,Bo:Diss2012}
\begin{IEEEeqnarray}{rCl}
V_S
& = & \sigma_U^2 \int_0^T e^{A(T-\tau)} b b^\T (e^{A(T-\tau)})^\T\, d\tau 
            \label{eqn:IntegratedVariance1}\IEEEeqnarraynumspace\\
& = & \sigma_U^2 \int_0^T e^{A\tau} b b^\T (e^{A\tau})^\T\, d\tau.
      \label{eqn:IntegratedVariance}
\end{IEEEeqnarray}
It is thus clear that,
for fixed $X(t_0)=x(t_0)$,
$X(t_1)$ is a Gaussian random vector with mean $e^{AT}x(t_0)$
and covariance matrix $V_S$,
i.e., 
\begin{equation} \label{eqn:fGaussian}
f\big(x(t_1)\cond x(t_0)\big) 
\propto e^{-\frac{1}{2}\left( x(t_1) - e^{AT}x(t_0) \right)^\T V_S^{-1} \left( x(t_1) - e^{AT}x(t_0) \right)}.
\end{equation}

\section{Factor Graph of System Model}
\label{sec:FG}

We will use Forney factor graphs (also known as normal factor graphs \cite{Forney:nr2001})
as in \cite{Lg:ifg2004} and \cite{LDHKLK:fgsp2007}. 
The nodes\andor{}boxes in such a factor graph represent factors 
and
the edges in the graph represent variables. 

\begin{figure}
\begin{center}
\setlength{\unitlength}{0.1mm}
\begin{picture}(690,550)(0,0)

  \put(130,470){\pos{br}{$X(t_{k})$}}
  \put(40,455){\vector(1,0){110}}
  \put(150,430){\framebox(50,50){$=$}}
  \put(200,455){\vector(1,0){110}}

  \put(310,415){\framebox(80,80){}}
  \put(350,530){\cent{$f\big( x(t_{k+1}) \cond x(t_k) \big)$}}

  \put(390,455){\vector(1,0){110}}
  \put(500,430){\framebox(50,50){$=$}}
  \put(550,455){\vector(1,0){120}}
  \put(565,470){\pos{bl}{$X(t_{k+1})$}}

  \put(175,430){\vector(0,-1){100}}
  \put(525,430){\vector(0,-1){100}}

  \put(135,250){\framebox(80,80){$C$}}
  \put(485,250){\framebox(80,80){$C$}}

  \put(175,250){\vector(0,-1){100}}
  \put(185,200){\pos{cl}{$Y_k$}}
  \put(525,250){\vector(0,-1){100}}
  \put(535,200){\pos{cl}{$Y_{k+1}$}}

  \put(0,100){\framebox(50,50){}}
  \put(25,85){\pos{ct}{$\mathcal{N}\!\left(0, V_Z\right)$}}
  \put(50,125){\vector(1,0){100}}
  \put(100,135){\pos{cb}{$Z_k$}}
  \put(150,100){\framebox(50,50){$+$}}
  \put(350,100){\framebox(50,50){}}
  \put(400,125){\vector(1,0){100}}
  \put(450,135){\pos{cb}{$Z_{k+1}$}}
  \put(500,100){\framebox(50,50){$+$}}

  \put(185,45){\pos{cl}{$\tilde{Y}_k = \tilde{y}_k$}}
  \put(175,100){\vector(0,-1){100}}
  \put(535,45){\pos{cl}{$\tilde{Y}_{k+1} = \tilde{y}_{k+1}$}}
  \put(525,100){\vector(0,-1){100}}

  \put(10,350){\cent{\ldots}}
  \put(700,350){\cent{\ldots}}

\end{picture}
\caption{\label{fig:SystemModelFactorGraph}%
Factor graph of the system model
with observations $\tilde{Y}_k = \tilde{y}_k$.}
\end{center}
\end{figure}
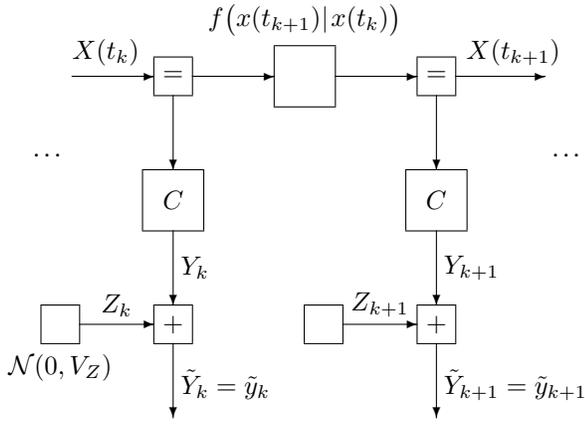

In this notation, the system model of Section~\ref{sec:SystemModel}
may be represented by the factor graph shown in \Fig{fig:SystemModelFactorGraph}. 
More precisely, \Fig{fig:SystemModelFactorGraph}
represents the joint probability density of the variables 
in the system model at discrete times $t_1,t_2,\ldots$. 

Note that \Fig{fig:SystemModelFactorGraph} 
shows only a section (from $t_k$ to $t_{k+1}$) of the factor graph; 
the complete factor graph starts at time $t_0$ and ends at some time $t_K$,
and it may contain additional nodes
to represent any pertinent initial or final conditions.
Note also that, apart from the Gaussian nodes\andor{}factors 
$f\big(x(t_{k+1}) \cond x(t_k)\big)$ and $\mathcal{N}(0,V_Z)$, 
the nodes\andor{}boxes in \Fig{fig:SystemModelFactorGraph} represent linear constraints.

For details of this factor graph notation, we refer to \cite{LDHKLK:fgsp2007}.

\begin{figure}
\setlength{\unitlength}{0.95mm}
\centering
\begin{picture}(65,32.5)(0,0)
%
\put(38,22.5){\framebox(5,5){}}    \put(37,25){\pos{cr}{$\mathcal{N}(0,V_S)$}}
\put(40.5,22.5){\vector(0,-1){12}}
\put(0,8){\vector(1,0){20}}        \put(7.5,9){\pos{bc}{$X(t_0)$}}
\put(20,4){\framebox(8,8){$e^{AT}$}}
\put(28,8){\vector(1,0){10}}
\put(38,5.5){\framebox(5,5){$+$}}
\put(43,8){\vector(1,0){22}}       \put(57.5,9){\pos{bc}{$X(t_1)$}}
\put(15,0){\dashbox(35,32.5)}
\end{picture}
\caption{\label{fig:SummarizedFactorGraph}%
Factor graph of $f\big(x(t_1)\cond x(t_0)\big)$
according to (\ref{eqn:Integrated})--(\ref{eqn:IntegratedVariance}).
(The values of $t_0$ and $t_1$ are not restricted to the discrete times $\{t_k\}$ 
in \Fig{fig:SystemModelFactorGraph}.)
}
\end{figure}
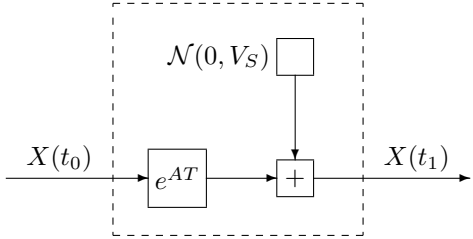

As shown in \Fig{fig:SystemModelFactorGraph},
the function (\ref{eqn:fGaussian}) can immediately 
be used as a node in a factor graph. 
However, the function (\ref{eqn:fGaussian}) 
can itself be represented by nontrivial factor graphs.
A first such factor graph is shown in \Fig{fig:SummarizedFactorGraph},
which corresponds to (\ref{eqn:Integrated})--(\ref{eqn:IntegratedVariance}). 
Plugging \Fig{fig:SummarizedFactorGraph} into 
\Fig{fig:SystemModelFactorGraph} results in a standard discrete-time
linear Gaussian factor graph as discussed in depth in \cite{LDHKLK:fgsp2007}.

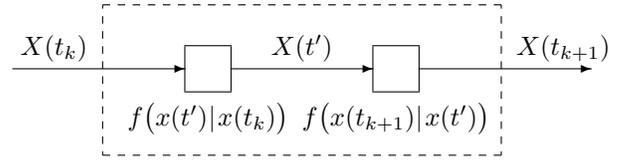
\begin{figure}
\setlength{\unitlength}{1mm}
\centering
\begin{picture}(53,20)(0,-1.5)
 \put(0,-1.5){\dashbox(53,20){}}
 \put(-12,10){\vector(1,0){23}} \put(-2,11){\pos{br}{$X(t_k)$}}
 \put(11,7){\framebox(6,6){}}   \put(14,5.5){\pos{ct}{$f\big( x(t') \cond x(t_k) \big)$}}
 \put(17,10){\vector(1,0){19}}  \put(26.5,11){\pos{cb}{$X(t')$}}
 \put(36,7){\framebox(6,6){}}   \put(39,5.5){\pos{ct}{$f\big( x(t_{k+1}) \cond x(t') \big)$}}
 \put(42,10){\vector(1,0){23}}  \put(55,11){\pos{bl}{$X(t_{k+1})$}}
\end{picture}
\caption{\label{fig:SplitNode}%
Splitting the node\andor{}factor $f\big( x(t_{k+1}) \cond x(t_k) \big)$
to access the state at an intermediate point in time~$t'$.}
\end{figure}

The factor graph of \Fig{fig:SystemModelFactorGraph} 
is easily refined to arbitrary temporal resolution 
by splitting the node\andor{}factor $f\big( x(t_{k+1}) \cond x(t_k) \big)$ 
as shown in \Fig{fig:SplitNode}. 
In this way, both the state $X(t)$ and the output signal $Y(t) = CX(t)$ 
become available for arbitrary instants $t$ between $t_k$ and $t_{k+1}$. 

Each of the factors in \Fig{fig:SplitNode} can, of course, 
be replaced by the corresponding decomposition 
according to \Fig{fig:SummarizedFactorGraph}.

Note that the input signal $U(t)$ is not explicitly represented 
in Figures \ref{fig:SystemModelFactorGraph}--\ref{fig:SplitNode}. 
For the estimation of $U(t)$, we will therefore need another 
decomposition of the node\andor{}factor $f\big( x(t_{k+1}) \cond x(t_k) \big)$.

\section{Gaussian Message Passing, State Estimation, and Output Signal Estimation}
\label{sec:GaussMessPass}

Having thus obtained a discrete-time factor graph 
(with an arbitrary temporal resolution), 
estimating $X(t)$ or $Y(t)$ 
from the noisy observations 
\mbox{$\tilde{Y}_1=\tilde{y}_1$}, \mbox{$\tilde{Y}_2=\tilde{y}_2$}, \ldots\
by means of Gaussian message passing 
is standard and discussed in detail in \cite{LDHKLK:fgsp2007} 
(cf.\ also \cite{Lg:ifg2004} and \cite{Wym:ird}). 
We therefore confine ourselves to a few general remarks 
(mostly excerpted from \cite{Lg:ifg2004} and \cite{LDHKLK:fgsp2007})
and some additional remarks on message passing through the 
node\andor{}factor $f\big( x(t) \cond x(t_0) \big)$.

\subsection{General Remarks}

\begin{enumerate}
\item
In linear Gaussian factor graphs such as 
Figures \mbox{\ref{fig:SystemModelFactorGraph}--\ref{fig:SplitNode}}
(where all factors are either Gaussians or linear constraints),
all sum-product messages are Gaussians and
sum-product message passing coincides with max-product message passing.
Moreover, MAP (maximum \emph{a posteriori}) estimation coincides both with 
MMSE (minimum mean squared error) estimation and with 
LMMSE (linear\andor{}affine
MMSE) estimation.
\item
In general, every edge in the factor graph carries two messages, 
one in each direction.
Since all the edges in Figures \ref{fig:SystemModelFactorGraph}--\ref{fig:SplitNode} 
are directed (i.e., drawn with an arrow), 
we can unambiguously refer to the forward message $\msgf{\mu}{X}$ 
and the backward message $\msgb{\mu}{X}$ along the edge representing some variable~$X$.
\item
Gaussian messages have the form 
\begin{equation}
\mu(x) \propto e^{-\frac{1}{2}(x-m)^\T W (x-m)};
\end{equation}
they are naturally parameterized 
by the mean vector $m$ and either the matrix $W$ 
or the covariance matrix $V$ ($=W^{-1}$).
Degenerate Gaussians, where either $W$ or $V$ do not have full rank, are often permitted
and sometimes unavoidable; in such cases, only $W$ or $V$, but not both, are well defined.
We will use the symbols $\msgf{m}{X}$ and $\msgf{V}{X}$ (or $\msgf{W}{X}$) 
to denote the parameters
of the forward message (along some edge\andor{}variable $X$) 
and $\msgb{m}{X}$ and $\msgb{V}{X}$ (or $\msgb{W}{X}$) 
for the parameters of the backward message.
\item
The natural scheduling 
of the message computations 
in \Fig{fig:SystemModelFactorGraph} 
consists of 
a forward recursion for $\msgf{\mu}{X(t_k)}$ 
and an independent backward recursion%
\footnote{The backward recursion is required for smoothing, i.e., noncausal estimation
 or estimation with delay; it is absent in basic Kalman filtering as in \cite{Ka:1960}.
 In fact, while the backward recursion is obvious from the graphical-model perspective, 
 its development in the traditional approach was not quite so obvious, cf.\ \cite{KSH:LinEs2000b}.}
for $\msgb{\mu}{X(t_k)}$. 
Both of these recursions use the messages $\msgb{\mu}{Y_k}$
with parameters $\msgb{m}{Y_k} = \tilde{y}_k$
and $\msgb{W}{Y_k} = V_z^{-1}$ 
(assuming $\tilde{Y}_k=\tilde{y}_k$ is known; 
if $\tilde{Y}_k$ is not observed\andor{}unknown, 
then $\msgb{W}{Y_k}=0$ and $\msgb{\mu}{Y_k}(y_k)=1$).
\item
Since the factor graph in \Fig{fig:SystemModelFactorGraph}
has no cycles, 
the \emph{a~posteriori} distribution of any variable $X$ (or $Y$, $Z$, \ldots) in the factor graph
is the product $\msgf{\mu}{X}(x) \msgb{\mu}{X}(x)$
of the corresponding two messages, up to a scale factor.
The parameters of this marginal distribution are $m_X$ and $W_X$ given by
$W_X = \msgf{W}{X} + \msgb{W}{X}$
and 
$W_X m_X = \msgf{W}{X}\msgf{m}{X} + \msgb{W}{X}\msgb{m}{X}$.
\item
Tabulated message computation rules 
(in particular, Tables 2--4 of \cite{LDHKLK:fgsp2007})
allow to compose a variety of different algorithms to compute the same sum-product messages. 
The variety arises from different parameterizations of the messages
and from local manipulations (including splitting and grouping of nodes) 
of the factor graph. 
\end{enumerate}

\subsection{Message Passing Through $f\big(x(t_1)\cond x(t_0)\big)$}
\label{sec:MessageComputationTable}

\begin{eqntable}
\caption{\label{tab:messagepassingrules}%
Computation rules for Gaussian messages
through node\andor{}factor 
$f\big(x(t_1)\cond x(t_0)\big)$ 
with $t_1 > t_0$.%
}
\begin{center}
\tablebox{
\begin{center}
\vspace{1mm}
\setlength{\unitlength}{0.1mm}
\begin{picture}(280,120)(0,0)
 %
 \put(-30,50){\makebox(100,50)[br]{$X(t_0)$}}
 \put(0,40){\vector(1,0){100}}
 \put(100,0){\framebox(80,80)[c]{}}
   \put(140,-30){\cent{$f\big(x(t_1)\cond x(t_0)\big)$}}
 \put(200,50){\makebox(100,50)[bl]{$X(t_1)$}}
 \put(180,40){\vector(1,0){100}}
\end{picture}
\end{center}
\vspace{6mm}
\begin{IEEEeqnarray}{rCl}
  \msgf{m}{X(t_1)} & = & e^{A(t_1-t_0)} \msgf{m}{X(t_0)}
  \label{TabEqn:msgfmX}\\
  \msgf{V}{X(t_1)} & = & e^{A(t_1-t_0)} \msgf{V}{X(t_0)} e^{A^\T (t_1-t_0)}
    \label{TabEqn:msgfVX}
  \nonumber\\
  & \relphantom{=}{} & 
    {} + \sigma_U^2 \underbrace{\int_0^{t_1-t_0} e^{A\tau} b b^\T e^{A^\T\tau} d\tau}_{Q\msgf{\Theta}{}(t_1-t_0)Q^\H\makebox[0em]{\hspace{5em}see~(\ref{eqn:IntF})}}
    \label{eqn:msgfV}\\[1.5ex]
  \msgb{m}{X(t_0)} & = & e^{-A(t_1-t_0)}\msgb{m}{X(t_1)}
     \label{TabEqn:msgbmX}\\
  \msgb{V}{X(t_0)} & = & e^{-A(t_1-t_0)} \msgb{V}{X(t_1)} e^{-A^\T (t_1-t_0)}
    \label{TabEqn:msgbVX}
  \nonumber\\ 
  & \relphantom{=}{} & 
    {} + \sigma_U^2 \underbrace{\int_0^{t_1-t_0} e^{-A \tau} bb^\T e^{-A^\T \tau} d\tau}_{Q\msgb{\Theta}{}(t_1-t_0)Q^\H\makebox[0em]{\hspace{5em}see~(\ref{eqn:IntB})}}
     \label{eqn:msgbV}\\[1.5ex]
  %
  \hat{u}(t)
      &=& \sigma_U^2 b^\T 
      \left(\msgf{V}{X(t)} + \msgb{V}{X(t)}\right)^{-1} 
      \left(\msgb{m}{X(t)} - \msgf{m}{X(t)}\right) 
      \label{TabEqn:mU}
      \nonumber \\
  &&
  %
\end{IEEEeqnarray}
\vspace{-2mm}
}
\end{center}
\end{eqntable}

Gaussian message passing through the node\andor{}factor $f\big(x(t_1)\cond x(t_0)\big)$
is summarized in Table~\ref{tab:messagepassingrules}.
Both the forward message 
(with parameters (\ref{TabEqn:msgfmX}) and (\ref{TabEqn:msgfVX}))
and the backward message 
(with parameters (\ref{TabEqn:msgbmX}) and~(\ref{TabEqn:msgbVX})) 
are easily obtained from \Fig{fig:SummarizedFactorGraph},
(\ref{eqn:IntegratedVariance1}) and (\ref{eqn:IntegratedVariance}),
and Tables 2 and~3 of \cite{LDHKLK:fgsp2007}. 

If the matrix $A$ is diagonalizable,  
then the integrals in (\ref{TabEqn:msgfVX}) and (\ref{TabEqn:msgbVX}) 
can easily be expressed in closed form.
Specifically, if 
\begin{equation} \label{eqn:diagA}
  A =  Q \left(\begin{array}{ccc}
    \lambda_1 &  & 0\\
    & \ddots & \\
    0 & & \lambda_n
  \end{array}\right) Q^{-1}
\end{equation}
for some complex square matrix $Q$, then
\begin{equation} \label{eqn:IntF}
\int_0^{t} e^{A\tau} b b^\T e^{A^\T\tau} d\tau
= Q \msgf{\Theta}{}(t) Q^\H
\end{equation}
where the square matrix $\msgf{\Theta}{}(t)$ is given by
\begin{equation} \label{eqn:ThetaF}
\msgf{\Theta}{}(t)_{k,\ell}
\eqdef \frac{(Q^{-1}b)_k \ccj{(Q^{-1}b)_\ell}}{\lambda_k + \ccj{\lambda_\ell}} \left( e^{(\lambda_k + \ccj{\lambda_\ell}) t} -1 \right),
\end{equation}
and
\begin{equation} \label{eqn:IntB}
\int_0^{t} e^{-A\tau} b b^\T e^{-A^\T\tau} d\tau
= Q \msgb{\Theta}{}(t) Q^\H
\end{equation}
with
\begin{equation} \label{eqn:ThetaB}
\msgb{\Theta}{}(t)_{k,\ell}
\eqdef \frac{(Q^{-1}b)_k \ccj{(Q^{-1}b)_\ell}}{\lambda_k + \ccj{\lambda_\ell}} \left( 1 - e^{-(\lambda_k + \ccj{\lambda_\ell}) t} \right).
\end{equation}
Note that, in (\ref{eqn:ThetaF}) and (\ref{eqn:ThetaB}), 
$(Q^{-1}b)_k$ denotes the $k$-th component of the vector $Q^{-1}b$.
The proof of (\ref{eqn:IntF}) and (\ref{eqn:IntB}) is given 
in Appendix~\ref{appsec:ProofIntF}.

The remaining entry (\ref{TabEqn:mU}) 
in Table~\ref{tab:messagepassingrules} 
is Theorem~\ref{theorem:InputEstimation} of the next section.

\section{Input Signal Estimation and Regularized-Least-Squares Interpretation}
\label{sec:InputEstimation}

\begin{figure}
\setlength{\unitlength}{0.95mm}
\centering
\begin{picture}(88,53)(0,0)
%
\put(30,36.5){\framebox(5,5){}}
  \put(36,42.5){\pos{br}{$\mathcal{N}\!\left( 0,\frac{\sigma_U^2 N}{T} \right)$}}
\put(32.5,36.5){\vector(0,-1){10}}
  \put(31.5,31.5){\pos{rc}{$\tilde{U}(t_0+\frac{T}{N})$}}
\put(28.5,18.5){\framebox(8,8){$b\frac{T}{N}$}}
\put(32.5,18.5){\vector(0,-1){8}}
\put(68,36.5){\framebox(5,5){}}
  \put(74,42.5){\pos{br}{$\mathcal{N}\!\left( 0,\frac{\sigma_U^2 N}{T} \right)$}}
\put(70.5,36.5){\vector(0,-1){10}}
  \put(69.5,31.5){\pos{rc}{$\tilde{U}(t_1)$}}
\put(66.5,18.5){\framebox(8,8){$b\frac{T}{N}$}}
\put(70.5,18.5){\vector(0,-1){8}}
\put(0,8){\vector(1,0){15}}        \put(9,9){\pos{br}{$X(t_0)$}}
\put(15,4){\framebox(8,8){$e^{\frac{AT}{N}}$}}
\put(23,8){\vector(1,0){7}}
\put(30,5.5){\framebox(5,5){$+$}}
\put(35,8){\line(1,0){5}}
\put(44,8){\pos{c}{$\ldots$}}
\put(48,8){\vector(1,0){5}}
\put(53,4){\framebox(8,8){$e^{\frac{AT}{N}}$}}
\put(61,8){\vector(1,0){7}}
\put(68,5.5){\framebox(5,5){$+$}}
\put(73,8){\vector(1,0){15}}        \put(79,9){\pos{lb}{$X(t_1)$}}
\put(10,0){\dashbox(68,53)}
\end{picture}
\caption{\label{fig:DiscreteFactorGraph}%
Decomposition of the node\andor{}factor $f\big(x(t_1)\cond x(t_0)\big)$
into $N$ discrete time steps (with $T= t_1-t_0$).
This representation is exact only in the limit $N\rightarrow\infty$.}
\end{figure}
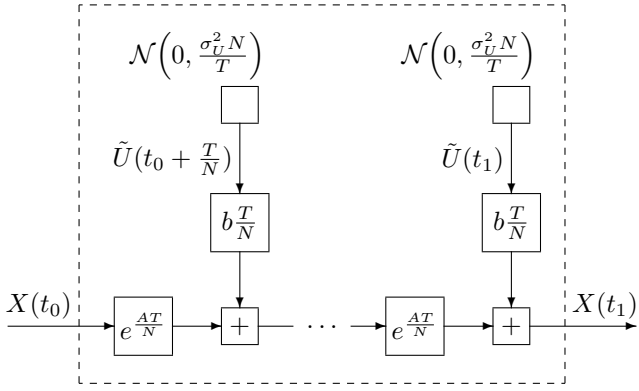

We now turn to estimating the input signal $U(t)$ and to clarifying its meaning.
To this end, we need the factor graph representation of $f\big(x(t_1)\cond x(t_0)\big)$
that is shown in \Fig{fig:DiscreteFactorGraph}, 
which corresponds to the decomposition of (\ref{eqn:Integrated}) into 
$N$ discrete steps and where $T \eqdef t_1-t_0$.
Note that this factor graph is only an approximate 
representation of $f\big(x(t_1)\cond x(t_0)\big)$, 
but the representation becomes exact in the limit $N\rightarrow\infty$. 
The variables $\tilde{U}(t)$ in \Fig{fig:DiscreteFactorGraph} 
are related to $U(t)$ by
\begin{equation} \label{eqn:DiscreteTimeUt}
\tilde{U}(t) = \frac{N}{T} \int_{t-T/N}^{t} U(\tau)\, d\tau,
\end{equation}
i.e., $\tilde{U}(t)$ is the average of $U(t)$ over the corresponding interval.
The proof of this decomposition is given in Appendix~\ref{appsec:DiscreteDecomp}\@. 

For finite $N$, \Fig{fig:DiscreteFactorGraph} is a standard 
linear Gaussian factor graph in which snapshots $\tilde{U}(t)$ of $U(t)$ 
according to (\ref{eqn:DiscreteTimeUt}) appear explicitly
and can therefore be estimated by standard Gaussian message passing.
In the resulting expression for the estimate of $\tilde{U}(t)$, 
we can take the limit $N\rightarrow \infty$ and thus obtain an estimate of $U(t)$.

\begin{theorem}\label{theorem:InputEstimation}
The MAP\andor{}MMSE\andor{}LMMSE estimate of $U(t)$ 
from observations $\tilde{Y}_k = \tilde{y}_k$
according to the system model of Section~\ref{sec:SystemModel} is 
\begin{equation} \label{eqn:InputEstimationTheorem}
\hat{u}(t) = \sigma_U^2 b^\T 
      \left(\msgf{V}{X(t)} + \msgb{V}{X(t)}\right)^{-1} 
      \left(\msgb{m}{X(t)} - \msgf{m}{X(t)}\right)
\end{equation}
where $\msgf{m}{X(t)}$, $\msgf{V}{X(t)}$, and $\msgb{m}{X(t)}$, $\msgb{V}{X(t)}$
are the parameters of 
the Gaussian sum-product messages as discussed in Section~\ref{sec:GaussMessPass}.
\end{theorem}

\begin{_proof}
Consider the factor graph in \Fig{fig:ProofInputEstim},
which shows the relevant part of \Fig{fig:DiscreteFactorGraph}
with suitably named variables.
We determine the mean $m_{\tilde{U}(t)}$ 
and the variance $W_{\tilde{U}(t)}^{-1}$
of the \emph{a posteriori} distribution of $\tilde{U}(t)$ as follows.
From \cite[eq.~(54) and (III.5)]{LDHKLK:fgsp2007}, 
we have
\begin{IEEEeqnarray}{rCl}
W_{\tilde{U}(t)} & = & \msgf{W}{\tilde{U}(t)} + \msgb{W}{\tilde{U}(t)} \\
 & = &  \sigma_U^{-2} \frac{T}{N} + \left(\frac{T}{N}\right)^{\! 2}\! b^\T \msgb{W}{\tilde{U}'(t)} b.
        \label{eqn:ProofInputMeanWU}
\end{IEEEeqnarray}

\begin{figure}
\setlength{\unitlength}{1mm}
\centering
\begin{picture}(35,51)(0,-3)
%
\put(15,35){\framebox(5,5){}}
  \put(17.5,41){\pos{bc}{$\mathcal{N}\!\left( 0,\frac{\sigma_U^2 N}{T} \right)$}}
\put(17.5,35){\vector(0,-1){11}}   \put(16.5,29.5){\pos{rc}{$\tilde{U}(t)$}}
\put(13.5,16){\framebox(8,8){$b\frac{T}{N}$}}
\put(17.5,16){\vector(0,-1){11}}   \put(16.5,10.5){\pos{rc}{$\tilde{U}'(t)$}}
%
\put(0,2.5){\vector(1,0){15}}      \put(7,1){\pos{ct}{$X'(t)$}}
\put(15,0){\framebox(5,5){$+$}}
\put(20,2.5){\vector(1,0){15}}     \put(28,1){\pos{ct}{$X(t)$}}
%
\end{picture}
\caption{\label{fig:ProofInputEstim}%
Factor graph used in the proof
of Theorem~\ref{theorem:InputEstimation}.}
\end{figure}
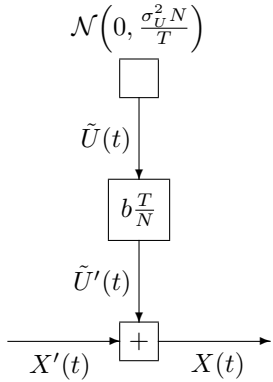

\noindent
From \cite[eq.~(55)]{LDHKLK:fgsp2007}, we then have 
\begin{equation}
W_{\tilde{U}(t)} m_{\tilde{U}(t)} 
  = \msgf{W}{\tilde{U}(t)} \msgf{m}{\tilde{U}(t)} + \msgb{W}{\tilde{U}(t)} \msgb{m}{\tilde{U}(t)};
\end{equation}
inserting $\msgf{m}{\tilde{U}(t)} = 0$ and using \cite[eq.~(III.6)]{LDHKLK:fgsp2007} yields
\begin{equation} \label{eqn:ProofInputMeanWUmU}
W_{\tilde{U}(t)} m_{\tilde{U}(t)} = \frac{T}{N} b^\T \msgb{W}{\tilde{U}'(t)} \msgb{m}{\tilde{U}'(t)}.
\end{equation}
Using (\ref{eqn:ProofInputMeanWU}) and (\ref{eqn:ProofInputMeanWUmU}), we obtain
\begin{IEEEeqnarray}{rCl}
m_{\tilde{U}(t)} & = & (W_{\tilde{U}(t)})^{-1} \left( W_{\tilde{U}(t)} m_{\tilde{U}(t)} \right) \\
 & = &  \left( \sigma_U^{-2} + \frac{T}{N} b^\T \msgb{W}{\tilde{U}'(t)} b \right)^{\!-1} \!
        b^\T \msgb{W}{\tilde{U}'(t)} \msgb{m}{\tilde{U}'(t)}
       \IEEEeqnarraynumspace \\
 & \approx & \sigma_U^2 b^\T \msgb{W}{\tilde{U}'(t)} \msgb{m}{\tilde{U}'(t)}
       \label{eqn:ProofInputMeanmU1}
\end{IEEEeqnarray}
and the approximation (\ref{eqn:ProofInputMeanmU1}) 
becomes exact in the limit \mbox{$N\rightarrow\infty$}.

Using \cite[eq.~(II.10)]{LDHKLK:fgsp2007}, we have
\begin{IEEEeqnarray}{rCl}
\msgb{m}{\tilde{U}'(t)} & = & \msgb{m}{X(t)} - \msgf{m}{X'(t)} \\
 & \approx &  \msgb{m}{X(t)} - \msgf{m}{X(t)},
              \label{eqn:ProofInputmU2}
\end{IEEEeqnarray}
and using \cite[eq.~(II.8)]{LDHKLK:fgsp2007}, we have
\begin{IEEEeqnarray}{rCl}
\msgb{W}{\tilde{U}'(t)} & = & \left( \msgb{V}{\tilde{U}'(t)} \right)^{-1} \\
 & = &  \left( \msgf{V}{X'(t)} + \msgb{V}{X(t)} \right)^{-1} \\
 & \approx & \left( \msgf{V}{X(t)} + \msgb{V}{X(t)} \right)^{-1}.
             \label{eqn:ProofInputWU2}
\end{IEEEeqnarray}
Again, the approximations (\ref{eqn:ProofInputmU2}) 
and (\ref{eqn:ProofInputWU2}) 
both become exact in the limit \mbox{$N\rightarrow\infty$}.
Inserting (\ref{eqn:ProofInputmU2}) and (\ref{eqn:ProofInputWU2}) 
into (\ref{eqn:ProofInputMeanmU1}) yields
\begin{equation} \label{eqn:ProofInputMean}
\lim_{N\rightarrow\infty} m_{\tilde{U}(t)} = \sigma_U^2 b^\T 
      \left(\msgf{V}{X(t)} + \msgb{V}{X(t)}\right)^{-1} 
      \left(\msgb{m}{X(t)} - \msgf{m}{X(t)}\right).
\end{equation}
The mean of the \emph{a posteriori} probability of $\tilde{U}(t)$ 
is thus well defined even for \mbox{$N\rightarrow\infty$} and given by (\ref{eqn:ProofInputMean}),
and the theorem follows.
\end{_proof}

While we have thus established that the mean (\ref{eqn:ProofInputMean}) 
of the \emph{a~posteriori} distribution of $\tilde{U}(t)$ is well-defined for $N\rightarrow\infty$,
it should be pointed out that the variance of this distribution is infinite:
taking the limit $N\rightarrow \infty$ of (\ref{eqn:ProofInputMeanWU})
yields $W_{\tilde{U}(t)}=0$. 
However, this seemingly problematic result does not imply 
that the estimate (\ref{eqn:InputEstimationTheorem}) is useless;
it simply reflects the obvious fact that white noise cannot be fully estimated 
from discrete noisy samples.

The nature of the estimate (\ref{eqn:InputEstimationTheorem})
is elucidated by the following theorem,
which reformulates the estimation problem of this paper
as an equivalent regularized least-squares problem.
For the sake of clarity,
we here restrict ourselves to scalar observations $Y_k$
where $\nu = 1$, $c\eqdef C$ is a row vector, and $\sigma_Z^2 \eqdef V_Z$ is a scalar.
(The general case is given in \cite{Bo:Diss2012}.)

\begin{theorem} \label{theorem:GlobalCost}
Assume that the factor graph in \Fig{fig:SystemModelFactorGraph}
consists of $K$ sections between $t_0$ and $t_K$
(with observations starting at $t_1$)
and assume that the observations $Y_k$ are scalars.
Then 
the estimated pair $\big(\hat{u}(t), \hat{x}(t)\big)$
with $\hat{u}(t)$ as in (\ref{eqn:InputEstimationTheorem})
minimizes 
\begin{equation} \label{eqn:GlobalCost}
\frac{1}{\sigma_U^2} \int_{t_0}^{t_K} \! \hat{u}(t)^2\, dt 
+ \frac{1}{\sigma_Z^2} \sum_{k=1}^K \big(\tilde{y}_k - c\hat{x}(t_k)\big)^2
\end{equation}
subject to the constraints of the system model. 
\end{theorem}

\begin{_proof}
Recall the factor graph representation of a least squares problem 
as in \Fig{fig:ProofGlobalCost}, where the large box on top 
expresses the given constraints. 
Clearly, maximizing the function represented by \Fig{fig:ProofGlobalCost}
amounts to computing
\begin{equation}  \label{eqn:CostFunctionGeneral}
\argmax_{z_1,\ldots,z_n} \prod_{k=1}^n e^{-z_k^2/(2\sigma_k^2)}
=  \argmin_{z_1,\ldots,z_n} \sum_{k=1}^n z_k^2/\sigma_k^2
\end{equation}
subject to the constraints. 
The right-hand side of (\ref{eqn:CostFunctionGeneral}) will be called ``cost function.''
Recall that sum-product message passing 
in cycle-free linear Gaussian factor graphs 
maximizes the left-hand side of (\ref{eqn:CostFunctionGeneral}) 
(subject to the constraints)
and thus minimizes the cost function \cite{LDHKLK:fgsp2007}.

\begin{figure}
\setlength{\unitlength}{1mm}
\centering
\begin{picture}(60,30)(0,0)
%
\put(0,15){\framebox(60,15){constraints}}
\put(7.5,5){\line(0,1){10}}   \put(6.5,10){\pos{cr}{$Z_1$}}
\put(5,0){\framebox(5,5){}}   \put(3.5,2.5){\pos{cr}{$\mathcal{N}(0,\sigma_1^2)$}}
\put(22.5,5){\line(0,1){10}}  \put(21.5,10){\pos{cr}{$Z_2$}}
\put(20,0){\framebox(5,5){}}
\put(37.5,4){\pos{c}{\ldots}}
\put(52.5,5){\line(0,1){10}}  \put(51.5,10){\pos{cr}{$Z_n$}}
\put(50,0){\framebox(5,5){}}  \put(56.5,2.5){\pos{cl}{$\mathcal{N}(0,\sigma_n^2)$}}
\end{picture}
\caption{\label{fig:ProofGlobalCost}%
Factor graphs of a least squares problem
used in the proof of Theorem~\ref{theorem:GlobalCost}.}
\end{figure}
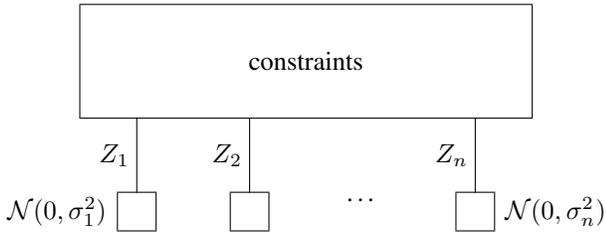

Now plugging \Fig{fig:DiscreteFactorGraph}
into the factor graph in \Fig{fig:SystemModelFactorGraph}
results in a factor graph as in \Fig{fig:ProofGlobalCost} 
with cost function

\begin{IEEEeqnarray}{rCl}
\IEEEeqnarraymulticol{3}{l}{
\sum_{k=1}^K \left( z_k^2/\sigma_Z^2 +  \sum_{\ell=1}^N 
              \tilde{u}\Big(t_{k-1}+\ell\frac{T_k}{N}\Big)^2 \frac{T_k}{\sigma_U^2 N} \right)
              \hspace{3em}
 }\nonumber\\\quad
 & = & \sum_{k=1}^K \left( z_k^2/\sigma_Z^2 + \frac{1}{\sigma_U^2} \int_{t_{k-1}}^{t_k} \!
              u(t)^2\, dt \right),
       \IEEEeqnarraynumspace
\end{IEEEeqnarray}
which is~(\ref{eqn:GlobalCost}).
\end{_proof}

According to Theorem~\ref{theorem:GlobalCost}, 
minimizing (\ref{eqn:GlobalCost}) is mathematically equivalent to 
the statistical estimation problems of this paper; 
in particular, modeling $U(t)$ as white Gaussian noise 
amounts to regularizing the second term in (\ref{eqn:GlobalCost})
by penalizing power in $\hat{u}(t)$.

The functional (\ref{eqn:GlobalCost}) is amenable
to an informal frequency-domain analysis 
that considers 
the relative power in the different frequences of the input signal $\hat{u}(t)$.
In particular, 
the estimate $\hat{u}(t)$ fits the corresponding output signal $\hat{y}(t) = c\hat{x}(t)$ 
to the observations $\tilde{y}_k$ preferably by those frequencies 
that appear with little damping in the output signal. 
Since the transfer function from $U(t)$ to $Y(t)=cX(t)$
of the system (\ref{eqn:ContSystDiffEq}) is necessarily a (non-ideal) low-pass filter,
the estimate $\hat{u}(t)$ will contain little energy in very high frequencies.
In this way, 
the spectrum of $\hat{u}(t)$ is shaped by the transfer function of the linear system.

We also note\footnote{%
This was pointed out to the authors by Andrew Singer of the University of Illinois at Urbana-Champain.}
that the problem of minimizing (\ref{eqn:GlobalCost}) 
may be viewed as 
an offline control problem
where an input signal $u(t)$ is to be determined
such that the resulting sampled output signal $y_1, y_2,\ldots$
follows a desired trajectory $\tilde{y}_1, \tilde{y}_2, \ldots$. 
However, exploring this connection to control theory is beyond the scope of this paper.

\section{Numerical Examples}
\label{sec:NumericalExamples}

We illustrate 
the estimators of this paper
by some simple numerical examples.
In all these examples, 
the output signal $Y(t)$ is scalar, 
we use regular sampling at rate $f_s$, i.e., $Y_k = Y(k/f_s)$,
and the linear system in \Fig{fig:SystemModel}
is a Butterworth lowpass filter of order 4 or~6
with cut-off frequency (\mbox{-3\,dB} frequency) $f_c$ \cite{OW:ss}.
The amplitude response (i.e., the magnitude of the frequency response)
of these filters is plotted in \Fig{fig:ExampleAmplitudeResponse}.

In these examples, we use the signal-to-noise ratio (SNR) 
as discussed in Appendix~\ref{appsec:SNR}.
Using (\ref{eqn:EY2Diag}), 
the SNR 
of the discrete-time observations turns out to be
\begin{equation}
\text{SNR} \approx \frac{\sigma_U^2}{\sigma_Z^2} f_c \cdot 2.052
\end{equation}
for the 4th-order filter and
\begin{equation}
\text{SNR} \approx \frac{\sigma_U^2}{\sigma_Z^2} f_c \cdot 2.023
\end{equation}
for the 6th-order filter. 
We will measure the SNR in dB (i.e., $10\!\cdot\!\log_{10}(\text{SNR})$).

In some of these plots, 
the estimator deliberately assumes an incorrect SNR, 
i.e., an incorrect ratio $\sigma_U^2/\sigma_Z^2$, 
in order to illustrate the effect of this ratio on (\ref{eqn:GlobalCost}).

\begin{figure}
\centering
\includegraphics{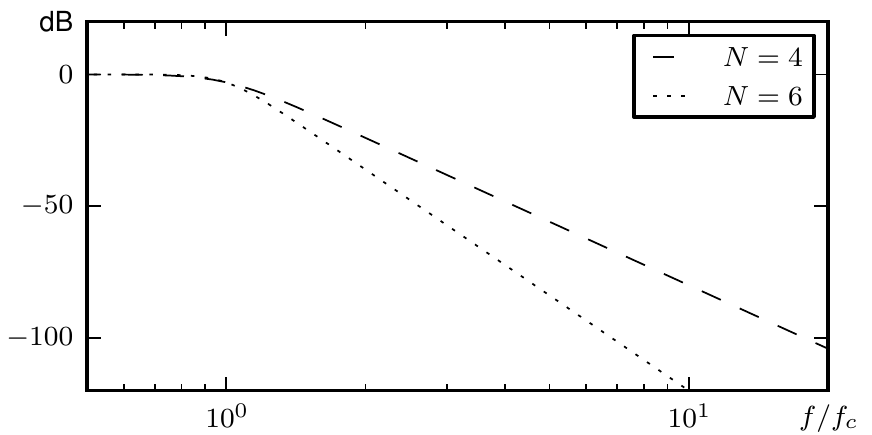}
\caption{\label{fig:ExampleAmplitudeResponse}%
Frequency response (magnitude) 
of the filters used in Section~\ref{sec:NumericalExamples}.}
\end{figure}

\begin{figure}
\centering
\includegraphics{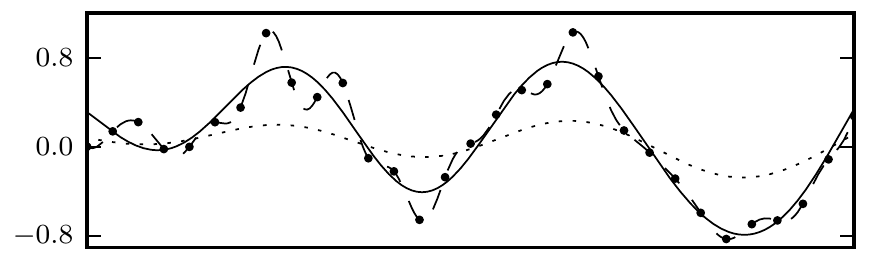}
\caption{\label{fig:plotoutest}%
Estimation of output signal $Y(t)$ 
from noisy samples $\tilde{y}_k$ (fat dots) at SNR = 10~dB.
Solid line: estimate of $Y(t)$ at correct SNR.
Dashed line: estimation with assumed SNR 100~dB;
dotted line: estimation with assumed SNR -10~dB.}
\end{figure}

Estimation of the output signal $Y(t)$ is illustrated in \Fig{fig:plotoutest}.
In this example, the linear system is a Butterworth filter of order~6. 
The noisy samples $\tilde{y}_k$ are created 
with $f_s = 10 f_c$ at an SNR of 10~dB.
The corresponding estimate of $Y(t)$ is shown as solid line in \Fig{fig:plotoutest}.

Also shown in \Fig{fig:plotoutest} is the effect
of estimating with an incorrect SNR, 
i.e., of playing with the ratio $\sigma_U^2/\sigma_Z^2$ as mentioned above.
Estimating with an assumed SNR that is too high results in overfitting;
estimating with an assumed SNR that is too low reduces the amplitude of the estimated signal.

\begin{figure}
\centering
\vspace{-7mm}
\includegraphics{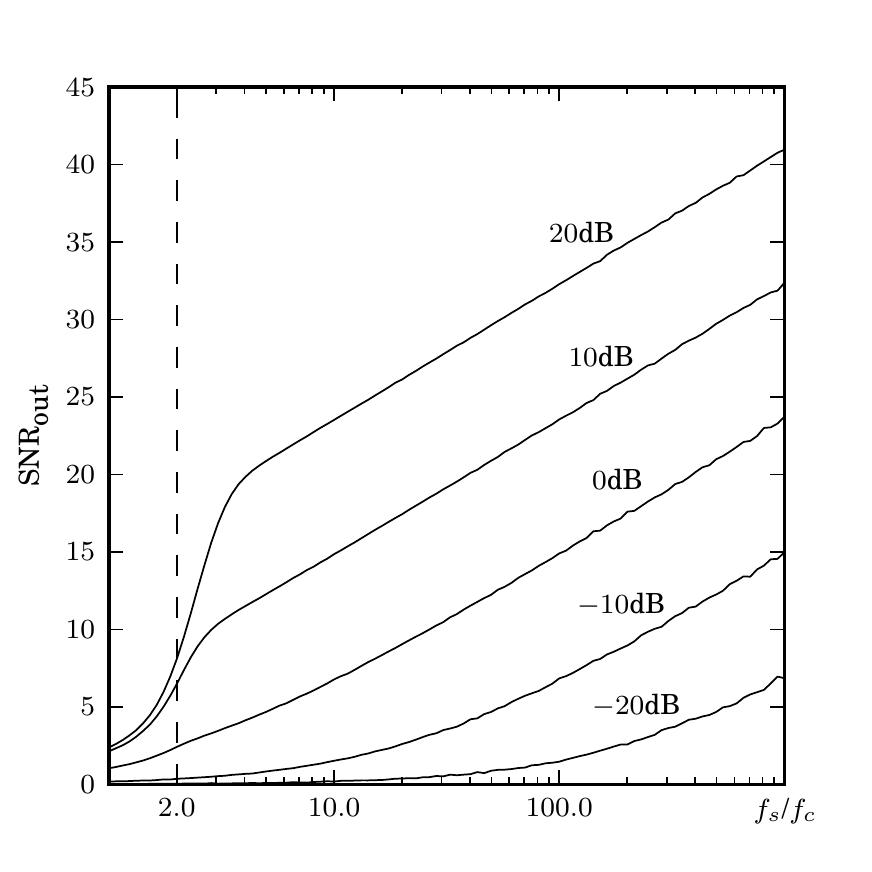}%
\vspace{-5mm}%
\caption{\label{fig:Oversampling}%
Empirical estimation error (\ref{eqn:NormEstError}) vs.\ normalized sampling frequency $f_s/f_c$, 
parameterized by the SNR (\ref{eqn:SNR}), 
for a Butterworth filter of order~4.%
}
\end{figure}

\Fig{fig:Oversampling} shows the effect of $f_s/f_c$ on 
the normalized estimation error
\begin{equation} \label{eqn:NormEstError}
\text{SNR}_\text{out}^{-1}
\eqdef  \frac{\EE{(\hat{Y}_k-Y_k)^2}}{\EE{Y_k^2}}
\end{equation}
for a Butterworth filter of order~4. 
For high SNR, we clearly see a critical ``Nyquist region'' 
where severe undersampling sets in.
For large $f_s/f_c$, 
the estimate improves by about 2.62~dB with every factor of 2 in $f_s/f_c$,
which is less than 
what would be expected (viz., 3~dB) 
for strictly bandlimited signals \cite{Be:sqs1948,TaVe:lbosq1996}.

\begin{figure}
  \begin{center}
    \includegraphics{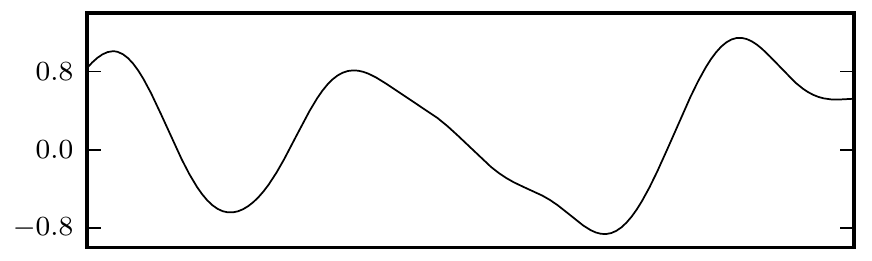}\\
    \includegraphics{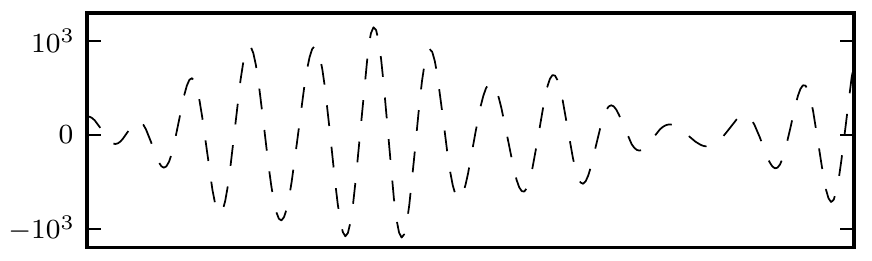}\\
    \includegraphics{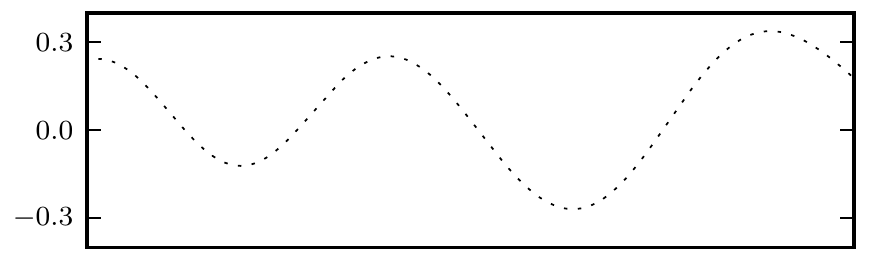}
 \caption{\label{fig:plotinpest}%
    Input signal estimation for the same cases (and the same time scale)
    as in \Fig{fig:plotoutest}. The solid line (top) is the correct MMSE\andor{}LMMSE estimate of $U(t)$.}
  \end{center}
\end{figure}

Estimation of the input signal $U(t)$ is illustrated in \Fig{fig:plotinpest}, 
for exactly the same setting (with the same discrete-time observations $\tilde{y}_k$)
as in \Fig{fig:plotoutest}. 
The power and the spectral content for the three different plots in \Fig{fig:plotinpest}
illustrate the effect of the ratio $\sigma_U^2/\sigma_Z^2$
on (\ref{eqn:GlobalCost}).

\section{Extensions}
\label{sec:Extensions}

We briefly mention a number of extensions and modifications 
of the system model that are required in some of the motivating applications
and are easily incorporated in the estimation algorithms.

\subsection{Additional Spectral Shaping}

The estimate (\ref{eqn:InputEstimationTheorem}) 
of the input signal $U(t)$
is marked by an implicit spectral shaping 
(cf.\ the discussion after Theorem~\ref{theorem:GlobalCost}).

It may sometimes be desirable, however, to control the spectrum 
of the estimate more explicitly.
This can be achieved by assuming that the input signal $U(t)$
is not white Gaussian noise, but white Gaussian noise passed through 
a suitable (finite-dimensional) linear prefilter. 
The estimation of $U(t)$ is easily adapted to this case
by including the prefilter in the system model.

In contrast to unfiltered-input estimation as in Section~\ref{sec:InputEstimation},
estimation 
of a filtered input signal 
by means of Kalman filtering\andor{}smoothing
is standard.

\subsection{Time-Varying and Affine Systems}

In some applications, 
the dynamics of the system\andor{}filter in \Fig{fig:SystemModel}
may change at discrete instants in time (but it is always known). 
This situation occurs, e.g., when the analog system\andor{}filter 
is subject to digital control. An example of such a case 
is given in \cite{LBWB:ITA2011c}.

We thus generalize the system model 
(\ref{eqn:ContSystDiff}) and (\ref{eqn:Yk})
to
\begin{equation} \label{eqn:TimeVarSystem}
dX(t) = \big( A_k X(t) + b_k U(t) + h_k \big)\, dt
\end{equation}
and
\begin{equation}
Y_k = C_k X(t_k),
\end{equation}
which holds for $t_k \leq t < t_{k+1}$, 
where $A_k$ and $C_k$ are known matrices, 
and where $b_k$ and $h_k$ are known column vectors. 

If $h_k=0$, 
both the factor graph representations and the message computation rules 
remain unchanged 
except for the addition of subscripts to the involved matrices and vectors.
The case $h_k\neq 0$ 
is included below.

\subsection{Multiple Inputs and Internal Noise}

We are also interested in the case where the system\andor{}filter in \Fig{fig:SystemModel}
has internal noise sources. 
(Again, a main motivation are analog-to-digital converters, 
where the noise in the analog part cannot be neglected.)
Such internal noise
can be handled mathematically by extending the input signal $U(t)$
to a vector $U(t) = \big( U_1(t), U_2(t), \ldots \big)^{\!\T}$,
where the first component, $U_1(t)$, is the actual input signal
while the remaining components model the internal noise.
For $t<t'$, the integral 
\begin{equation}
\int_t^{t'} \! U(\tau)\, d\tau
\end{equation}
is a zero-mean Gaussian random vector with diagonal covariance matrix 
$\sigma_U^2 I (t'-t)$. 
The corresponding generalization of (\ref{eqn:ContSystDiff}) is
\begin{equation} \label{eqn:TimeVarSystemMultInput}
d X(t) = \big( A X(t) + B U(t) + h\big)\, dt,
\end{equation}
where $B$ is a matrix of suitable dimensions and where we have included 
a constant offset $h$ (a column vector) as in (\ref{eqn:TimeVarSystem}). 
Note that power differences and correlations among the 
input signals can be expressed by a suitable matrix $B$.

The corresponding generalization of Table~\ref{tab:messagepassingrules}
is shown in Table~\ref{tab:GenMessageRules}. 
The proofs are straightforward modifications 
of the proofs of Table~\ref{tab:messagepassingrules} and 
are omitted.

\begin{eqntable}
\caption{\label{tab:GenMessageRules}%
Generalization of Table~\ref{tab:messagepassingrules}
to (\ref{eqn:TimeVarSystemMultInput}).
}
\begin{center}
\tablebox{
\begin{center}
\vspace{1mm}
\setlength{\unitlength}{0.1mm}
\begin{picture}(280,120)(0,0)
 %
 \put(-30,50){\makebox(100,50)[br]{$X(t_0)$}}
 \put(0,40){\vector(1,0){100}}
 \put(100,0){\framebox(80,80)[c]{}}
   \put(140,-30){\cent{$f\big(x(t_1)\cond x(t_0)\big)$}}
 \put(200,50){\makebox(100,50)[bl]{$X(t_1)$}}
 \put(180,40){\vector(1,0){100}}
\end{picture}
\end{center}
\vspace{6mm}
\begin{IEEEeqnarray}{rCl}
  \msgf{m}{X(t_1)} & = & e^{A(t_1-t_0)} \msgf{m}{X(t_0)}
  + A^{-1} \big( e^{A(t_1-t_0)} - I \big) h
  \IEEEeqnarraynumspace
  \label{TabEqnGen:msgfmX}\\
  \msgf{V}{X(t_1)} & = & e^{A(t_1-t_0)} \msgf{V}{X(t_0)} e^{A^\T (t_1-t_0)}
    \label{TabEqnGen:msgfVX}
  \nonumber\\
  & \relphantom{=}{} & 
    {} + \sigma_U^2 
    \underbrace{\int_0^{t_1-t_0} e^{A\tau} B B^\T e^{A^\T\tau} d\tau}_{Q\msgf{\Theta}{}(t_1-t_0)Q^\H\makebox[0em]{\hspace{5em}see~(\ref{eqn:GenThetaF})}}
    \label{TabEqnGen:msgfV}\\[1.5ex]
  %
  \msgb{m}{X(t_0)} & = & e^{-A(t_1-t_0)} 
     \nonumber\\
     & \relphantom{=}{} & 
     \Big( \msgb{m}{X(t_1)} - A^{-1} \big( e^{A(t_1-t_0)} - I \big) h \Big)
      \IEEEeqnarraynumspace
      \label{TabEqnGen:msgbmX}\\
  \msgb{V}{X(t_0)} & = & e^{-A(t_1-t_0)} \msgb{V}{X(t_1)} e^{-A^\T (t_1-t_0)}
    \label{TabEqnGen:msgbVX}
  \nonumber\\ 
  & \relphantom{=}{} & 
    {} + \sigma_U^2 
    \underbrace{\int_0^{t_1-t_0} e^{-A \tau} B B^\T e^{-A^\T \tau} d\tau}_{Q\msgb{\Theta}{}(t_1-t_0)Q^\H\makebox[0em]{\hspace{5em}see~(\ref{eqn:GenThetaB})}}
     \label{TabEqnGen:msgbV}\\[1.5ex]
  %
  \hat{u}(t)
      &=& \sigma_U^2 B^\T 
      \left(\msgf{V}{X(t)} + \msgb{V}{X(t)}\right)^{-1} 
      \left(\msgb{m}{X(t)} - \msgf{m}{X(t)}\right) 
      \label{TabEqnGen:mU}
      \nonumber \\
  &&
  %
\end{IEEEeqnarray}
\vspace{-2mm}
}
\end{center}
\end{eqntable}

If the matrix $A$ is diagonalizable as in (\ref{eqn:diagA}),
then the integrals in (\ref{TabEqnGen:msgfV}) and (\ref{TabEqnGen:msgbV})
can be written as stated in the table 
where the square matrices $\msgf{\Theta}{}(t)$ and $\msgb{\Theta}{}(t)$ 
are given by
\begin{align} 
  \msgf{\Theta}{}(t)_{k,\ell}
  \eqdef \frac{\psi_{k,\ell}}{\lambda_k + \ccj{\lambda_\ell}} 
  \left( e^{(\lambda_k + \ccj{\lambda_\ell}) t} -1 \right)
  \label{eqn:GenThetaF}
\end{align}
and by
\begin{align} 
  \msgb{\Theta}{}(t)_{k,\ell}
  \eqdef \frac{\psi_{k,\ell}}{\lambda_k + \ccj{\lambda_\ell}} 
  \left( 1 - e^{-(\lambda_k + \ccj{\lambda_\ell}) t} \right),
  \label{eqn:GenThetaB}
\end{align}
respectively,
and
where $\psi_{k,\ell}$ is the entry in row $k$ and column $\ell$ of the matrix
\begin{equation}
\Psi \eqdef Q^{-1}B\, (Q^{-1}B)^\H.
\end{equation}

\subsection{Nonlinearities}

Mild nonlinearities in the system\andor{}filter in \Fig{fig:SystemModel}
can often be handled
by extended Kalman filtering \cite{GrAn:KF,KSH:LinEs2000b}, 
i.e., by iterative estimation using a linearized model based on a tentative estimate 
of the state trajectory $X(t)$.

\section{Conclusions}
\label{sec:Conclusion}

We have pointed out that exact models of continuous-time linear systems 
driven by white Gaussian noise can be used in discrete-time factor graphs. 
The associated continuous-time signals then become 
computational objects that can be handled with arbitrary 
temporal resolution by discrete-time Gaussian message passing. 

Motivated by applications such as dynamical sensors and analog-to-digital converters,
we have been particularly interested in estimating the input signal,
which does not seem to have been addressed in the prior Kalman filtering literature.

\appendices

\section{Proof of (\ref{eqn:IntF}) and (\ref{eqn:IntB})}
\label{appsec:ProofIntF}

Let
\begin{equation}
\Lambda \eqdef 
  \left(\begin{array}{ccc}
    \lambda_1 &  & 0\\
    & \ddots & \\
    0 & & \lambda_n
  \end{array}\right).
\end{equation}
From (\ref{eqn:diagA}), we have
\begin{equation} \label{eqn:expA_diag}
e^{A\tau} = Q e^{\Lambda\tau} Q^{-1}
\end{equation}
and
\begin{equation}
e^{A^\T\!\tau} = (e^{A\tau})^\T = (e^{A\tau})^\H = (Q^{-1})^\H e^{\ccj{\Lambda}\tau} Q^\H,
\end{equation}
and thus
\begin{equation}
\int_0^{t} e^{A\tau} b b^\T e^{A^\T\tau} d\tau
 =  Q \left( \int_0^{t} e^{\Lambda\tau} \Psi e^{\ccj{\Lambda}\tau} \, d\tau \right) Q^\H
\end{equation}
with
\begin{equation}
\Psi \eqdef Q^{-1}b (Q^{-1}b)^\H.
\end{equation}
The element in row $k$ and column $\ell$ of the matrix under the integral is
\begin{equation}
\left( e^{\Lambda\tau} \Psi e^{\ccj{\Lambda}\tau} \right)_{k,\ell}
= \psi_{k,\ell}\, e^{(\lambda_k + \ccj{\lambda_\ell}) \tau},
\end{equation}
where $\psi_{k,\ell}$ refers to the elements of the matrix $\Psi$,
and elementwise integration yields
\begin{equation}
\left( \int_0^{t} e^{\Lambda\tau} \Psi e^{\ccj{\Lambda}\tau} \, d\tau \right)_{k,\ell}
=  \frac{\psi_{k,\ell}}{\lambda_k + \ccj{\lambda_\ell}} \left( e^{(\lambda_k + \ccj{\lambda_\ell}) t} -1 \right),
\end{equation}
which proves (\ref{eqn:IntF}).
The proof of (\ref{eqn:IntB}) follows from 
noting that changing $e^{A\tau}$ into $e^{-A\tau}$
amounts to a sign change of $\Lambda$.

\section{Proof of the Discrete-Time Decomposition in \Fig{fig:DiscreteFactorGraph}}
\label{appsec:DiscreteDecomp}

\begin{figure}
\setlength{\unitlength}{0.95mm}
\centering
\begin{picture}(88,67)(0,-14)
%
\put(14.5,36.5){\framebox(5,5){}}
  \put(13.5,42.5){\pos{bl}{$\mathcal{N}\!\left( 0,\frac{\sigma_U^2 T}{N} \right)$}}
\put(17,36.5){\vector(0,-1){10}}
  \put(18,31.5){\pos{lc}{$\tilde{U}(t_0+\frac{T}{N})$}}
\put(13,18.5){\framebox(8,8){$b\frac{T}{N}$}}
\put(17,18.5){\line(0,-1){10.5}}
\put(17,8){\vector(1,0){4}}
\put(35,36.5){\framebox(5,5){}}
\put(37.5,36.5){\vector(0,-1){10}}
\put(33.5,18.5){\framebox(8,8){$b\frac{T}{N}$}}
\put(37.5,18.5){\vector(0,-1){8}}
\put(68,36.5){\framebox(5,5){}}
  \put(74,42.5){\pos{br}{$\mathcal{N}\!\left( 0,\frac{\sigma_U^2 T}{N} \right)$}}
\put(70.5,36.5){\vector(0,-1){10}}
  \put(69.5,31.5){\pos{rc}{$\tilde{U}(t_1)$}}
\put(66.5,18.5){\framebox(8,8){$b\frac{T}{N}$}}
\put(70.5,18.5){\vector(0,-1){8}}
\put(21,4){\framebox(8,8){$e^{\frac{AT}{N}}$}}
\put(29,8){\vector(1,0){6}}
\put(35,5.5){\framebox(5,5){$+$}}
\put(40,8){\line(1,0){2}}
\put(45.5,8){\pos{c}{$\ldots$}}
\put(49,8){\vector(1,0){5}}
\put(54,4){\framebox(8,8){$e^{\frac{AT}{N}}$}}
\put(62,8){\vector(1,0){6}}
\put(68,5.5){\framebox(5,5){$+$}}
\put(70.5,5.5){\vector(0,-1){9}}
\put(0,-6){\vector(1,0){40}}        \put(9,-5){\pos{br}{$X(t_0)$}}
\put(40,-10){\framebox(8,8){$e^{AT}$}}
\put(48,-6){\vector(1,0){20}}
\put(68,-8.5){\framebox(5,5){$+$}}
\put(73,-6){\vector(1,0){15}}        \put(79,-5){\pos{lb}{$X(t_1)$}}
\put(10,-14){\dashbox(68,67)}
\end{picture}
\caption{\label{fig:DiscreteFactorGraphProof}%
Decomposition of the node\andor{}factor $f\big(x(t_1)\cond x(t_0)\big)$
into $N$ discrete time steps according to~(\ref{eqn:DiscreteFactorGraphProved}).}
\end{figure}
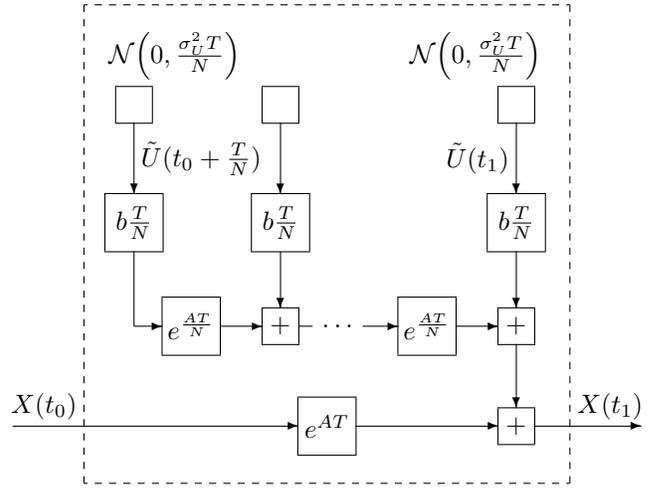

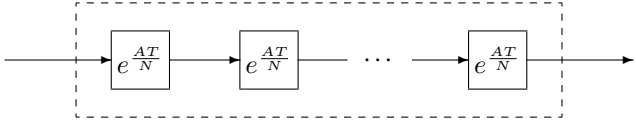
\begin{figure}
\setlength{\unitlength}{0.95mm}
\centering
\begin{picture}(88,16)(0,0)
%
\put(0,8){\vector(1,0){15}}
\put(15,4){\framebox(8,8){$e^{\frac{AT}{N}}$}}
\put(23,8){\vector(1,0){10}}
\put(33,4){\framebox(8,8){$e^{\frac{AT}{N}}$}}
\put(41,8){\line(1,0){7}}
 \put(52.5,8){\cent{\ldots}}
\put(57,8){\vector(1,0){8}}
\put(65,4){\framebox(8,8){$e^{\frac{AT}{N}}$}}
\put(73,8){\vector(1,0){15}}

\put(10,0){\dashbox(68,16)}
\end{picture}
\caption{\label{fig:DecompExpAT}%
Decomposition of $e^{AT}$ into $N$ sections.}
\end{figure}

We split the integral (\ref{eqn:Integrated}) into $N$ parts, each of width $T/N$
with $T\eqdef t_1-t_0$:
\begin{IEEEeqnarray}{rCl}
X(t_1)
 & = & e^{AT} X(t_0) \nonumber\\
     && {}\hspace{1em} + \sum_{k=1}^N \int_{(k-1)T/N}^{kT/N} \! e^{A(T-\tau)} bU(t_0+\tau)\, d\tau
             \IEEEeqnarraynumspace\\
 & \approx & e^{AT} X(t_0) \nonumber\\
     && {} + \sum_{k=1}^N e^{A(T-kT/N)} b \int_{(k-1)T/N}^{kT/N} \! U(t_0+\tau)\, d\tau
             \IEEEeqnarraynumspace\label{eqn:IntApprox} \\
 & = & e^{AT} X(t_0) \nonumber\\
     && {}\hspace{1em} + \sum_{k=1}^N e^{A(T-kT/N)} b \frac{T}{N} \tilde{U}(t_0+kT/N),
           \IEEEeqnarraynumspace\label{eqn:DiscreteFactorGraphProved}
\end{IEEEeqnarray}
where the approximation (\ref{eqn:IntApprox}) becomes exact 
in the limit \mbox{$N\rightarrow\infty$}
and where $\tilde{U}(t)$ is defined as in (\ref{eqn:DiscreteTimeUt}).
The factor graph of (\ref{eqn:DiscreteFactorGraphProved}) is shown in 
\Fig{fig:DiscreteFactorGraphProof}.

The term $e^{AT}$ can also be decomposed into $N$ discrete steps 
as shown in \Fig{fig:DecompExpAT}. 
Plugging \Fig{fig:DecompExpAT} into \Fig{fig:DiscreteFactorGraphProof} 
yields a factor graph that is easily seen to be equivalent to \Fig{fig:DiscreteFactorGraph}.

\section{On SNR}
\label{appsec:SNR}

For the system model of Section~\ref{sec:SystemModel},
we may wish to relate the input noise power $\sigma_U^2$ to the signal-to-noise ratio (SNR)
of the discrete-time observations. 
For the sake of clarity, we restrict ourselves to scalar observations $Y_k$, 
i.e., $\nu=1$, $c\eqdef C$ is a row vector, and $\sigma_Z^2 \eqdef V_Z$ is a scalar. 
In addition, we assume that 
the continuous-time linear system is time-invariant and stable
and any initial conditions can be neglected. 
In this case, we define
\begin{equation} \label{eqn:SNR}
\text{SNR} \eqdef \frac{\EE{Y_k^2}}{\sigma_Z^2}
\end{equation}
which (under the stated assumptions) 
is independent of $k$. 
We then have
\begin{equation}
\EE{Y_k^2} = c \msgf{V}{\!X(\infty)} c^\T
\end{equation}
with
\begin{equation}
\msgf{V}{\!X(\infty)} \eqdef \sigma_U^2 \lim_{t\rightarrow \infty}
                     \int_0^{t} e^{A \tau} bb^\T e^{A^\T \tau} d\tau
\end{equation}
from (\ref{TabEqn:msgfVX}); 
if, in addition, the system is diagonalizable as in (\ref{eqn:diagA}), 
then
\begin{equation} \label{eqn:EY2Diag}
\EE{Y_k^2} = \sigma_U^2  cQ \msgf{\Theta}{}(\infty) Q^\H c^\T
\end{equation}
where $\msgf{\Theta}{}(\infty)$ is a square matrix with entries
\begin{equation}
\msgf{\Theta}{}(\infty)_{k,\ell}
= - \frac{(Q^{-1}b)_k \ccj{(Q^{-1}b)_\ell}}{\lambda_k + \ccj{\lambda_\ell}}
\end{equation}

\newcommand{\IT}{IEEE Trans.\ Inform.\ Theory}
\newcommand{\CASI}{IEEE Trans.\ Circuits \& Systems~I}
\newcommand{\COM}{IEEE Trans.\ Comm.}
\newcommand{\COMLet}{IEEE Commun.\ Lett.}
\newcommand{\COMMag}{IEEE Communications Mag.}
\newcommand{\ETT}{Europ.\ Trans.\ Telecomm.}
\newcommand{\SPMag}{IEEE Signal Proc.\ Mag.}
\newcommand{\ProcIEEE}{Proceedings of the IEEE}

\end{document}